\documentclass[preprint2]{aastex}

\usepackage{epsfig}
\usepackage[]{natbib}

\newcommand{\eg}{{\rm e.g.}}
\newcommand{\ie}{{\rm i.e.,}}

\newcommand{\asec}{\ensuremath{\arcsec}}

\newcommand{\kmsMpc}{~\ensuremath{{\rm km\ s}^{-1}\ {\rm Mpc}^{-1}}}

\newcommand{\hMpc}{~\ensuremath{{h}^{-1}\ {\rm Mpc}}}

\newcommand{\ergsHzMpccube}{~\ensuremath{{\rm ergs\ s}^{-1}\ {\rm Hz}^{-1}\ {\rm Mpc}^{-3}}}

\newcommand{\uprime}{\ensuremath{U^{\prime}}}
\newcommand{\hkprime}{\ensuremath{HK^{\prime}}}

\newcommand{\ang}{\AA}

\newcommand{\Hnought}{\ensuremath{{\rm H}_0}}
\newcommand{\znought}{\ensuremath{z_0}}

\def \figwidth {\linewidth}

\slugcomment{To appear in The Astronomical Journal}

\shorttitle{Title}
\shortauthors{Wilson \etal}

\begin{document}

\title{STAR FORMATION HISTORY SINCE $Z = 1.5$ AS INFERRED FROM REST-FRAME ULTRAVIOLET 
LUMINOSITY DENSITY EVOLUTION }

\author{Gillian Wilson\altaffilmark{1}, Lennox L. Cowie\altaffilmark{2}, 
Amy J. Barger\altaffilmark{2,3,4}, and D. J. Burke\altaffilmark{2,5}}

\altaffiltext{1}{Physics Department, Brown University, 182 Hope Street, 
Providence, RI 02912}
\altaffiltext{2}{Institute for Astronomy, University of Hawaii, 
2680 Woodlawn Drive, Honolulu, HI 96822}
\altaffiltext{3}{Department of Astronomy, University of Wisconsin-Madison,
475 North Charter Street, Madison, WI 53706}
\altaffiltext{4}{Hubble Fellow and Chandra Fellow at Large}
\altaffiltext{5}{Harvard-Smithsonian Center for Astrophysics, 60 Garden Street, Cambridge,
MA 
02138}

\begin{abstract}
We investigate the evolution of the universal rest-frame ultraviolet 
luminosity density from $z = 1.5$ to the present. We analyze an extensive
sample of multicolor data (\uprime$_{AB}$, $B_{AB}$, $V_{AB} = 24.5$) 
plus spectroscopic redshifts from the Hawaii Survey Fields and the 
Hubble Deep Field.
Our multicolor data allow us to select our sample in the rest-frame
ultraviolet (2500 \ang) over the entire redshift range to $z = 1.5$.
We conclude that the evolution in the luminosity density is a function
of the form $(1+z)^{1.7\pm1.0}$ for a flat lambda
($\Omega_{{\rm m}0} = 0.3, \Omega_{\lambda 0} = 0.7$) cosmology
and $(1+z)^{2.4\pm1.0}$ for an Einstein-de Sitter  cosmology. 

\end{abstract}


\section{INTRODUCTION}
\label{sec:intro}

A major goal of observational cosmology is to understand the
star formation history of the Universe from the earliest
epoch of structure formation to the present.
Much recent
attention has focused on determining the contribution to the global 
history from the most distant sources;
however, the star formation history even at modest redshifts 
($z<1$) is not well determined and has recently undergone a revision.

Early work by \markcite{mad-96}{Madau} {et~al.} (1996) (later updated by  \markcite{mad-98}Madau, Pozzetti, \& Dickinson 1998)
suggested that the global star formation as seen in the optical 
and ultraviolet (UV) had a strong peak around $z=1$ and then fell 
very steeply at lower redshifts. The $z<1$ data used in the analysis
were taken from a paper by
\markcite{lilly-96}Lilly {et~al.} (1996), who used rest-frame 
near-UV luminosities derived from the $I$-selected Canada-France
Redshift Survey (CFRS, \markcite{lilly-95}{Lilly} {et~al.} 1995)
to determine the comoving UV luminosity density from 
$z=1$ to the present. These authors found the evolution to be a steep 
function of the form $(1 + z)^{4}$.
However, when \markcite{treyer-98}{Treyer} {et~al.} (1998) presented 
the first UV-selected constraints on the local integrated luminosity 
density, they found that their result was well above the optically-derived 
estimates. \markcite{sull-00}{Sullivan} {et~al.} (2000) subsequently
tripled the \markcite{treyer-98}{Treyer} {et~al.} UV-selected sample and confirmed the higher local
volume-averaged star formation rate.

\markcite{csb-99}Cowie, Songaila, \& Barger (1999, hereafter CSB) decided to reinvestigate the rest-frame UV 
luminosity density evolution to $z=1$ using a large, extremely deep, 
and highly complete spectroscopic galaxy redshift survey.
Their data enabled them to select objects based on the rest-frame
UV magnitudes at all redshifts.
The evolution found by these authors was a much shallower 
function of the form $(1 + z)^{1.5}$.
CSB suggested that the differences between their results and those
of  \markcite{lilly-96}Lilly {et~al.} could be accounted for by the $I$-band selection
of the \markcite{lilly-96}Lilly {et~al.} sample,
which required large extrapolations to obtain 
UV colors, and by the CFRS data not being deep enough to probe the 
flat segments of the luminosity function (LF), which meant that at
redshifts near $z=1$, reliable extrapolations to total luminosity 
density could not be made.

In this paper, we expand on the work of CSB to more thoroughly 
investigate the rest-frame UV luminosity density 
evolution from $z = 1.5$ to the present. Our new galaxy sample is
nearly twice as large as that used by CSB, and we explore various
methods for constructing the UV LFs. 
The outline of the paper is as follows.
In \S~\ref{sec:obs} we present 
our data sample and strategy and we investigate the \uprime\ number 
counts and redshift distribution. We
also explore the redshift-magnitude relationship for the
\uprime, $B$, and $V$ passbands. In \S~\ref{sec:UVLF} we describe 
how we construct rest-frame UV LFs
as a function of redshift from the \uprime, $B$, and $V$ data. 
In \S~\ref{sec:ld} we utilize these LFs to infer the evolution of 
the global UV luminosity density with redshift.
In \S~\ref{sec:conc} we summarize our conclusions. 
Initially, we assume a flat lambda ($\Omega_{{\rm m}0} = 0.3, \Omega_{\lambda 0} = 0.7$) 
cosmology with $\Hnought = 100$ h \kmsMpc. Subsequently, we 
investigate the effect of an Einstein-de Sitter ($\Omega_{{\rm m}0} = 1.0, \Omega_{\lambda 0} = 0.0$) cosmology on our results.  

\section{OBSERVATIONS}
\label{sec:obs}

We analyzed a three passband subset (\uprime\ ($3400\pm 150$\ang), 
$B$, $V$) of an ongoing eight passband 
(\uprime, $B$, $V$, $R$, $I$, $Z$, $J$, \hkprime) 
Hawaii imaging survey of four $6' \times 2.5'$ areas 
crossing the Hawaii Survey Fields SSA13, SSA17, and SSA22
\markcite{lilly-91}({Lilly}, {Cowie}, \& {Gardner} 1991) and the 
Hubble Deep Field (HDF) \markcite{will-96}({Williams} {et~al.} 1996).
The $B$ and $V$ images were obtained using 
the Low-Resolution Imaging Spectrograph 
(LRIS; \markcite{oke-95}Oke {et~al.} 1995) on the Keck~10~m telescopes 
and the UH8K CCD Mosaic Camera \markcite{lup-97}(Luppino 1997)
on the Canada-France-Hawaii 3.6~m telescope. The \uprime\ data 
were taken with the ORBIT CCD on the University of Hawaii 2.2~m 
telescope. All magnitudes were measured in 3\asec\ diameter apertures 
and corrected to total magnitudes following the procedures described 
in \markcite{cghskw-94}Cowie {et~al.} (1994). 

The sources analyzed in this paper contain and extend the CSB catalogs.
The total number of galaxies with spectroscopic redshifts
in the present \uprime, $B$, and $V$ samples are 403, 414, and 518, 
all to a survey limit magnitude (AB) of $24.5$.
These numbers can be compared to those in CSB 
(218, 350, and 259 respectively) . Thus, our current dataset contains approximately 
double the number of objects in the CSB \uprime\ and $V$ samples.

The great advantage of multiband data is that one can, for all 
redshifts, select galaxies based on their rest-frame UV magnitudes, 
thereby avoiding the uncertainties associated with selecting galaxies 
at longer (redder) wavelengths and then extrapolating to obtain 
their UV magnitudes. In addition, the depth of the current dataset 
allows us to construct LFs to sufficiently faint absolute magnitudes 
to constrain the faint-end slope. Moreover, by selecting galaxies 
based on their rest-frame UV magnitudes, we
expect the relative shape of the inferred UV LFs to be minimally 
sensitive to the effect of interstellar dust; \ie\ we expect the 
relative forms of our derived LFs (and hence the relative
values of our luminosity densities) to be subject only to evolution 
in the amount or properties of galactic dust with redshift
and not to its effect in absolute terms. (Note that any extinction
due to dust in the source galaxy should not be confused with the
tiny extinction corrections due to interstellar dust)

As discussed in CSB, one is free to choose the rest-frame wavelength 
at which to compute LFs and the UV light density. 
For our dataset, 2500 \ang\  provides a sensible compromise between 
our wide range of redshifts and large number of galaxies
in each passband. 

%
%

\begin{figure}
\centering\epsfig{file=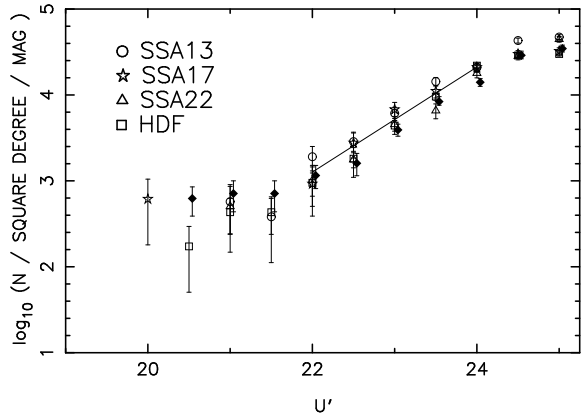,width=\figwidth}
\figcaption[plotcounts.ps]{
Number counts 
versus apparent magnitude for the \uprime\ sample. The four
fields are as indicated by the key.
The solid line shows the best-fit to the counts
between $U^{\prime}$ of 22.0 and 24.0
(slope of 0.61 and intercept at $-10.34$). The filled diamonds
are the counts from Hogg et al.\ (1997).
\label{fig:counts}
}
\end{figure}

Because there are missing or unidentified galaxies in each sample,
we construct LFs by either omitting them (\ie\ we assume
they are unidentified because they are unusual in some way, \eg,   
at very high redshift), or by assuming they are 
distributed in redshift in exactly the same manner as the identified 
galaxies (\ie\ we assume they are identical to the measured galaxies 
but have been missed for some trivial reason). It is likely that 
the true LF lies between these two possibilities. In subsequent 
sections we refer to the former case as minimal and the latter case as 
incompleteness-corrected. The \uprime, $B$, and $V$ samples are $88\%$, 
$90\%$, and $83\%$ complete, respectively, so the correction is not 
a huge factor in any case.  

\subsection{The $U^{\prime}$ Sample}
\label{ssec:uband}

Figure~\ref{fig:counts} shows number counts versus apparent magnitude
for our \uprime\ sample. The symbols indicate the number counts in
each of the four fields. We assume a 1 $\sigma$ Poisson uncertainty
for each field. The solid line shows the best-fit to the counts
for galaxies with magnitudes between \uprime\ of 22.0 and 24.0
(slope of $0.61\pm0.06$ and intercept at $-10.34\pm1.49$). The
uncertainties were calculated from the field-to-field variations.

A number of groups (e.g., 
\markcite{will-96,poz-98,gard-00}{Williams} {et~al.} 1996; Pozzetti {et~al.} 1998; {Gardner}, {Brown}, \& {Ferguson} 2000) 
have measured deep galaxy counts at 3000~\AA\ from the {\it HST} 
Hubble Deep Field imaging survey, but their counts are somewhat 
deeper than those presented here, so it is difficult to make a 
direct comparison. The sample 
most similar to ours was that obtained at the Palomar 5~m by 
\markcite{hogg-97}{Hogg} {et~al.} (1997). 
We overplot their data (filled diamonds) on Fig.~\ref{fig:counts}
for comparison, after converting their magnitudes to 
\uprime$_{AB}$ by adding 0.79~mag. The agreement is generally
good.

\begin{figure}
\centering\epsfig{file=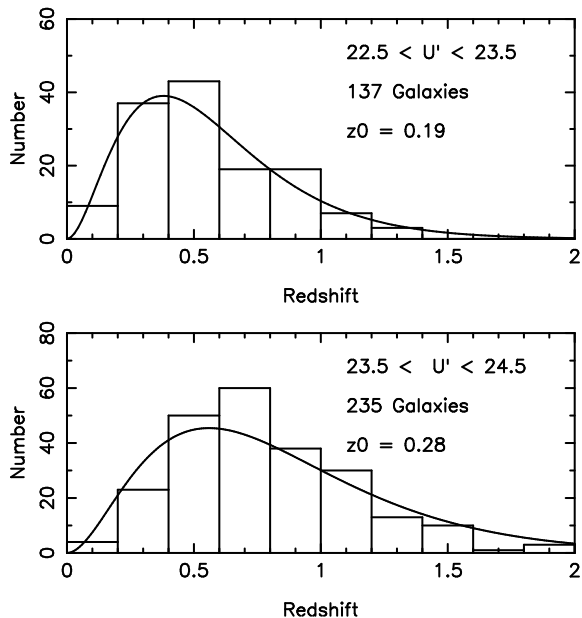,width=\figwidth}
\figcaption[plothist.ps]{
Redshift distributions for the \uprime\ sample. The upper panel
shows $N(z)$ for a one magnitude wide band 
($22.5<U^{\prime}\leq 23.5$) in apparent magnitude.
At this depth the data are $97\%$ complete. The solid line
shows the best-fit to the model (Eq.~\ref{eq:pz})
with redshift scale parameter 0.19. The lower panel
shows $N(z)$ for $23.5<U^{\prime}\leq 24.5$.
At this depth the data are $87\%$ complete and best-fit
with redshift scale parameter 0.28.
\label{fig:nz}
}
\end{figure}

In Fig.~\ref{fig:nz} we show the total number of objects in our 
\uprime\ sample versus redshift for two apparent magnitude bins.
The upper panel shows the number of galaxies versus
redshift for all galaxies with ${U^{\prime}}$ between
22.5 and 23.5, and the lower panel is for
galaxies with ${U^{\prime}}$ between 23.5 and 24.5.

%
%

 %
%

\begin{figure}
\centering\epsfig{file=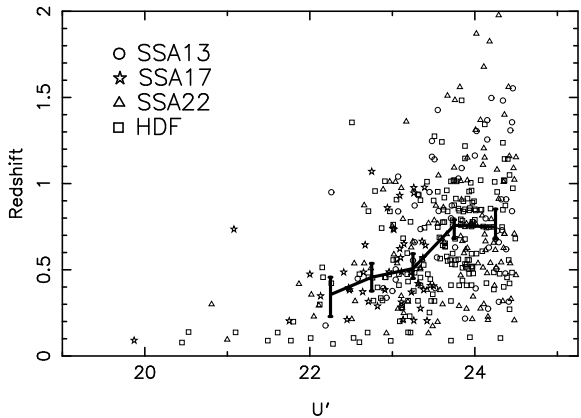,width=\figwidth}
\figcaption[plotmagz.ps]{
Galaxy redshift versus apparent magnitude for the
\uprime\ sample. The key is as in Fig.~\ref{fig:counts}.
The solid line shows the median redshift and the 1 $\sigma$ 
Poisson uncertainties (Table~\ref{tab:medz}).
\label{fig:zm}
}
\end{figure}

\begin{deluxetable}{ccrcrcr}
\tablewidth{0pt}
\tablecaption{Median redshifts and galaxy numbers as a function of apparent magnitude and wavelength (The uncertainties
are 1 $\sigma$ Poissonian limits). \label{tab:medz}}
\tablewidth{0pt}
\tablehead{
\colhead{Mag.}  & 
\colhead{$z_{\rm med}(U')$} &    \colhead{$N(U')$} &
\colhead{$z_{\rm med}(B)$} &    \colhead{$N(B)$} &
\colhead{$z_{\rm med}(V)$} &    \colhead{$N(V)$} 
}
\startdata
$22.25$  & $0.356$ $^{0.457}_{0.229}$ & $17$  & $0.380$ $^{0.452}_{0.317}$ & $44$  & $0.411$ $^{0.471}_{0.348}$ & $45$ \
\\
$22.75$  & $0.457$ $^{0.537}_{0.377}$ & $43$  & $0.475$ $^{0.560}_{0.432}$ & $55$  & $0.475$ $^{0.524}_{0.455}$ & $65$ \
\\
$23.25$  & $0.507$ $^{0.593}_{0.450}$ & $94$  & $0.663$ $^{0.777}_{0.564}$ & $86$  & $0.626$ $^{0.742}_{0.531}$ & $113$ \
\\
$23.75$  & $0.758$ $^{0.790}_{0.682}$ & $127$ & $0.715$ $^{0.788}_{0.617}$ & $112$ & $0.680$ $^{0.753}_{0.615}$ & $141$  \
\\
$24.25$  & $0.748$ $^{0.852}_{0.680}$ & $108$ & $0.753$ $^{1.010}_{0.562}$ & $77$  & $0.850$ $^{0.980}_{0.709}$ & $104$ \
\\
\enddata
\end{deluxetable}
 
\markcite{wklc-01}{Wilson} {et~al.} (2001) found 
that a good model for the redshift distribution (at least at 
$I$ and $V$) is provided by
\begin{equation}
\label{eq:pz}
p(z)  = 0.5 z^{2} \exp(-z/z_0) / {z_0}^{3}
\end{equation}
where $p(z)\times dz$ is the probability of finding a galaxy in 
the redshift interval $z+dz$ (the mean redshift is 
$\overline{z}=3 z_0$ and the
median redshift is $z_{\rm median} = 2.67 z_0$).
A nice property of (\ref{eq:pz}) is that there is only one free 
parameter, the redshift scale parameter, $z_0$.
The solid lines overlaid on Fig.~\ref{fig:nz}
show the best-fits to the model ($\znought=0.19$ and $\znought=0.28$), 
normalized to the total number of galaxies in each sample.

\subsection{Magnitude-Redshift Dependence}
\label{ssec:zm}

In this section we investigate the dependence of median redshift
on apparent magnitude and wavelength. In Fig.~\ref{fig:zm} we show 
redshift versus magnitude for the \uprime\ sample. 
The symbols denote the field in which the galaxy was observed. 
Note that the SSA13, SSA22, and HDF fields are $\sim90\%$ complete 
to an AB limiting magnitude
of 24.5, 
and the SSA17 field is similarly complete to 23.5.

Table~\ref{tab:medz} quantifies Fig.~\ref{fig:zm}, giving
the median redshift with $\pm$ 1 $\sigma$ Poisson
uncertainties \markcite{gehr-86}({Gehrels} 1986) and the number 
of objects in each half-magnitude interval as a function of apparent
magnitude. We note that the median redshift obtained from the parameterized  
fit to the data (Eq.~\ref{eq:pz}) for the $22.5<{U^{\prime}}\leq 23.5$ 
interval is 0.51, and for the $23.5<{U^{\prime}}\leq 24.5$ interval, 
0.75. The values are in good agreement with
those calculated directly in  Table~\ref{tab:medz}.

In Table~\ref{tab:medz} we also calculate the median redshift
for the $B$ and $V$ samples. Median redshift as a function of 
apparent magnitude can  potentially be used to constrain galaxy 
evolution models, \eg\ if we compare with the $B$-band predictions 
from the merger model proposed by \markcite{carl-92}{Carlberg} (1992, his Table~2), 
we find that his median redshift values of $0.44$ for 
$B=23$ and $0.55$ for $B=24$ are lower than our values, suggesting 
that more pure luminosity evolution might have occurred 
than was proposed in his model.

\section{THE REST-FRAME ULTRAVIOLET LUMINOSITY FUNCTION} 
\label{sec:UVLF}

For any given galaxy, $i$, at redshift, $z$, the equation relating the 
absolute and apparent magnitudes is given by

\begin{equation}
\label{eq:Mmz}
M_{i}^{2500} = m_{i} - 5 \log d_{L}(z_{i}) - 25  + 2.5 \log (1+z) + dK(z_{i})
\end{equation}

\noindent 
where $m$ is the observed magnitude at the redshifted wavelength and
$d_{L}(z)$ is the luminosity distance in \hMpc. 
$dK(z)$ is given by

\begin{equation}
\label{eq:dK}
dK(z)= 2.5 \log_{10} \frac{f_{\nu}[2500{\rm \AA}\times(1+z)]}{f_{\nu}[2500{\rm \AA}\times(1+z_{c})]}
\end{equation}

\noindent
where $f_{\nu}$ is the spectral energy distribution of the galaxy
and $z_{c}$ is the redshift corresponding to the center of the band.
$dK(z)$ is a differential $K$-correction to account for each sample containing a range of 
redshifts and hence a range of rest-frame wavelengths around 2500 \ang.
It is generally small and is obtained by interpolation 
from the neighboring passbands.

\subsection{ Luminosity Functions from the $V_{\rm max}$ Method}
\label{ssec:Vmax}

We used two methods to construct LFs:  the traditional $V_{\rm max}$ method 
described by \markcite{schmidt-68}Schmidt (1968), \markcite{fel-76}Felten (1976), and \markcite{ellis-96}Ellis {et~al.} (1996)
and a new method recently suggested by
\markcite{page-00}{Page} \& {Carrera} (2000). In the $V_{\rm max}$ method,
the number density of galaxies in the redshift range $[z_{1},z_{2}]$ with 
absolute magnitude $M$ is given by
 
\begin{equation}
\label{eq:VMAX}
\phi(M) \,dM  =  \sum\frac{1}{V_{\rm max}(M)}
\end{equation}

\noindent 
where the sum is over the galaxies in the magnitude interval $M\pm dM/2$.
$V_{\rm max}(M)$ is the maximum total comoving volume within which each galaxy
(as defined by its apparent magnitude and redshift) would remain detectable within
survey limits. The uncertainty for each magnitude interval is conventionally 
calculated from 

\begin{equation}
\label{eq:VMAXerr}
\sigma  =  \left[\sum\frac{1}{(V_{\rm max}(M))^{2}}\right]^\frac{1}{2}
\end{equation}

\noindent 
\markcite{marsh-85, boyle-88}({Marshall} 1985; {Boyle}, {Shanks}, \& {Peterson} 1988). 
This expression weights each observation
by its contribution to the sum. However, it assumes Gaussian statistics,
which is not ideal for bins at the bright or faint end of the LF 
where only a small number of objects contribute to the sum.

\begin{figure}
\centering\epsfig{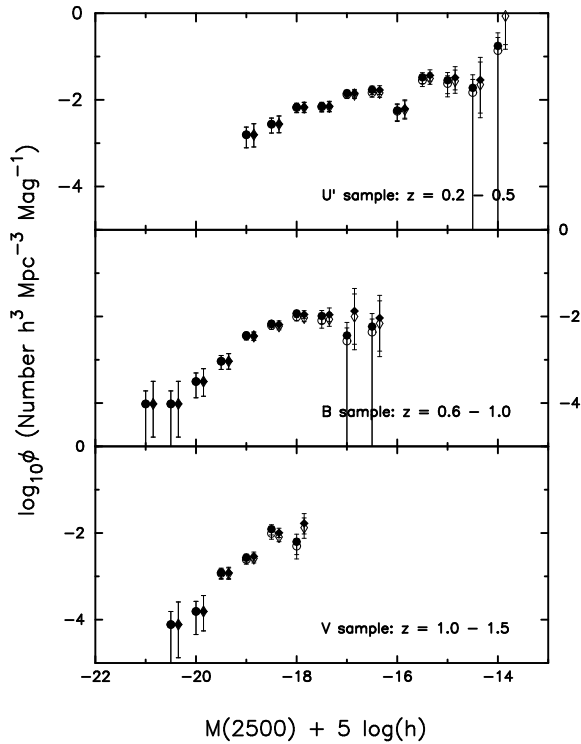}
\figcaption[paper_3plot_FL.ps]{
The 2500 \ang\ rest-frame luminosity function for three redshift 
bins: $z = 0.2-0.5$, from the \uprime\ sample (upper panel), 
$z = 0.6-1.0$, from the $B$ sample (center panel), and $z = 1.0-1.5$,
from the $V$ sample (lower panel). An
$\Omega_{{\rm m}0} = 0.3, \Omega_{\lambda 0} = 0.7$ cosmology is assumed.
The open circles denote the minimal function, and
the solid circles denote the incompleteness-corrected function
obtained using the $V_{\rm max}$ method.
The diamonds (offset by 0.15~mag for clarity) show the
minimal (open) and incompleteness-corrected (solid)
functions obtained using the $V_{\rm PC}$ method.
See text for details and an explanation of the uncertainties.
For this dataset the luminosity functions obtained from the two 
methods are very similar.
\label{fig:3plotmethod}
}
\end{figure}

Figure~\ref{fig:3plotmethod} shows the 2500 \ang\ rest-frame LF for our 
three redshift bins: $z = 0.2-0.5$,
constructed from the \uprime\ sample (upper panel), $z = 0.6-1.0$,
constructed from the $B$ sample (center panel), and $z = 1.0-1.5$,
constructed from the $V$ sample (lower panel). At redshifts $z = 0.35\pm0.15$,  
$0.80\pm0.20$ and $1.25\pm0.25$, the \uprime\ (3400\ang), B (4500\ang) and V (5500\ang) 
samples
correspond to rest-wavelengths of $2519^{+314}_{-253}$, $2500^{+312}_{-250}$ and 
$2444^{+306}_{-244}$ \ang. The open circles denote 
the minimal function, and the solid circles denote the 
incompleteness-corrected function. The number of galaxies used in the 
construction of each LF was 121 (\uprime), 119 ($B$), and 59 ($V$).

%
%

\subsection{ Luminosity Functions from the PC Method}
\label{ssec:PC}

The \markcite{page-00}{Page} \& {Carrera} (2000) method is very similar to 
the $V_{\rm max}(M)$ method in that it results in a binned differential LF,
but the advantage is that it more accurately determines the LF at the faint end.
The maximum redshift at which any galaxy may be found is a 
constantly varying function determined by the flux limit of the survey. The
$V_{\rm max}(M)$ method assumes that the redshift is a constant for any
given absolute magnitude bin. However, by dividing each magnitude bin into a series
of steps, calculating $V_{\rm max}(M)$ for each interval, and then integrating
over the magnitude bin, one can obtain a more precise estimate of the 
maximum volume to which each object could be observed in the survey.
Estimates of $\phi(M)$ obtained by either of the two methods should agree at the
bright end of the LF where objects are much brighter than the flux limit. 
The two estimates should also agree in the case of very fine magnitude 
intervals, such that the widths of the magnitude bins tend to zero.

We denote the LFs obtained using the $V_{\rm PC}$ method by diamonds
in Fig.~\ref{fig:3plotmethod}. The open triangles  show the minimal function and
the solid triangles show the incompleteness-corrected function.
The diamonds have been offset from the 
circles of the $V_{\rm max}$ method by 0.15 mag for clarity. 
The $\pm$ 1 $\sigma$ Poisson uncertainties 
\markcite{gehr-86}({Gehrels} 1986) in the LFs
are appropriate for small numbers of objects per magnitude interval.
The LFs obtained from the two methods are very similar.

Both $V_{\rm max}(M)$  and $V_{\rm PC}(M)$ give unbiased estimates of 
$\phi(M)$ only if galaxy clustering can be neglected. 
It is easy to imagine how a nearby excess of clustering could bias 
the estimator: such an excess of intrinsically faint
galaxies would cause the LF to be too steep at the faint end. 
Although the effects of clustering are mostly of concern in pencil-beam 
surveys, we tested two maximum likelihood alternatives, which 
should be less sensitive to clustering, on our \uprime\ sample. 
These were the Schechter fit estimator suggested by 
\markcite{sty-79}Sandage, Tammann, \& Yahil (1979)
and the stepwise estimator of 
\markcite{efs-88}Efstathiou, Ellis, \& Peterson (1988). 
In all cases, the LFs were very similar to those obtained 
using $V_{\rm max}$, giving us confidence in the robustness 
of our results. We also note that the LFs in this paper were obtained 
using code written independently from that used in CSB.

\subsection{Schechter Parameterization}
\label{ssec:param}

We then assumed that each LF could be parameterized by a Schechter function
\begin{equation}
\label{eq:Schechter}
\phi(M)\,dM=k\phi^{\star}
e^{k(\alpha+1)(M^{\star}-M)}
e^{-e^{k(M^{\star}-M)}}\,dM
\end{equation}

\noindent 
where $k=\frac{2}{5}\ln10$ \markcite{schecht-76}(Schechter 1976).
We solved for the best-fit Schechter parameters assuming two fixed 
faint-end slopes that
likely bound the range of faint-end slopes : 
$\alpha=-1.0$ and $\alpha=-1.5$. Table~{\ref{tab:bestfitFL}} shows the best-fit absolute magnitude
at the knee, $M^\star$, and normalization, $\phi^\star$ for these
values of fixed faint-end slope. In Fig.~\ref{fig:3plotcosmo} the open circles again 
show the minimal function; 
the solid circles the incompleteness-corrected function 
assuming a flat lambda cosmology 
as in 
Figure~\ref{fig:3plotmethod}.
Overlaid on Fig.~\ref{fig:3plotcosmo}
are the best-fitting Schechter functions assuming $\alpha=-1.0$ or $\alpha=-1.5$.
The solid portion of the line shows the magnitude range utilized
in the fit.

\begin{deluxetable}{ccccccccc}
\tablewidth{0pt}
\tablecaption{Schechter function parameter fits for minimal and incompleteness-corrected points for flat lambda   cosmology. The columns show the best-fit values of absolute magnitude
at the knee ($M^\star$), normalization ($\phi^\star$),  
and reduced $\chi^{2}$ for 
fixed faint-end slope ($\alpha$) of $-1.0$ and $-1.5$. 
\label{tab:bestfitFL}
}
\tablewidth{0pt}
\tablehead{
\colhead{}    &  \colhead{} & \multicolumn{3}{c}{$\alpha = -1.0$} &    \multicolumn{3}{c}{$\alpha = -1.5$} & \colhead{}\\
\colhead{Sample}  & \colhead{Corrected?} &  \colhead{$M^\star$} & \colhead{$\phi^\star$} & \colhead{$\chi^{2}/\nu$} & \colhead{$M^\star$} & \colhead{$\phi^\star$}  & \colhead{$\chi^{2}/\nu$} & \colhead{Range}  
}
\startdata
$U'$  & No  &  $-18.17$ & $0.0142$ & $1.90$ & $-19.86$ & $0.0024$ & $2.51$ & $[-19.25:-14.75]$ 
\\
$U'$  & Yes &  $-18.08$ & $0.0161$ & $1.25$ & $-19.64$ & $0.0031$ & $1.58$ & $[-19.25:-14.75]$ 
\\
$B$   & No  &  $-18.46$ & $0.0178$ & $0.55$ & $-18.96$ & $0.0089$ & $0.86$ & $[-21.25:-17.25]$ 
\\
$B$   & Yes &  $-18.34$ & $0.0232$ & $0.20$ & $-18.81$ & $0.0123$ & $0.36$ & $[-21.25:-17.25]$ 
\\
$V$   & No  &  $-18.30$ & $0.0212$ & $0.87$ & $-18.53$ & $0.0183$ & $0.75$ & $[-20.75:-18.25]$ 
\\
$V$   & Yes &  $-18.12$ & $0.0348$ & $0.69$ & $-18.35$ & $0.0306$ & $0.58$ & $[-20.75:-18.25]$ 
\\
\enddata
\end{deluxetable}

\subsection{Effect of Cosmology}
\label{ssec:cosmo} 

\begin{deluxetable}{ccccccccc}
\tablewidth{0pt}
\tablecaption{Schechter function parameter fits for minimal and incompleteness-corrected points for Einstein-de Sitter  cosmology. The columns show the best-fit values of absolute magnitude
at the knee ($M^\star$),  normalization ($\phi^\star$) and reduced $\chi^{2}$
for fixed faint-end slope ($\alpha$) of $-1.0$ and $-1.5$. 
\label{tab:bestfitEdS}
}
\tablewidth{0pt}
\tablehead{
\colhead{}    &  \colhead{} & \multicolumn{3}{c}{$\alpha = -1.0$} &    \multicolumn{3}{c}{$\alpha = -1.5$} & \colhead{}  \\
\colhead{Sample}  & \colhead{Corrected?} &  \colhead{$M^\star$} & \colhead{$\phi^\star$} & \colhead{$\chi^{2}/\nu$} & \colhead{$M^\star$} & \colhead{$\phi^\star$}  & \colhead{$\chi^{2}/\nu$} & \colhead{Range}  
}
\startdata
$U'$  & No  &  $-17.64$ & $0.0322$ & $0.41$ & $-19.02$ & $0.0070$ & $0.88$ & $[-18.75:-14.75]$
\\
$U'$  & Yes &  $-17.52$ & $0.0378$ & $0.22$ & $-18.75$ & $0.0094$ & $0.41$ & $[-18.75:-14.75]$ 
\\
$B$   & No  &  $-17.86$ & $0.0512$ & $0.26$ & $-18.23$ & $0.0318$ & $0.25$ & $[-20.75:-17.25]$ 
\\
$B$   & Yes &  $-17.75$ & $0.0643$ & $0.09$ & $-18.13$ & $0.0405$ & $0.05$ & $[-20.75:-17.25]$ 
\\
$V$   & No  &  $-17.63$ & $0.0812$ & $0.19$ & $-17.85$ & $0.0699$ & $0.21$ & $[-20.25:-17.75]$ 
\\
$V$   & Yes &  $-17.54$ & $0.1100$ & $0.04$ & $-17.78$ & $0.0932$ & $0.04$ & $[-20.25:-17.75]$ 
\\
\enddata
\end{deluxetable}

\begin{figure}
\centering\epsfig{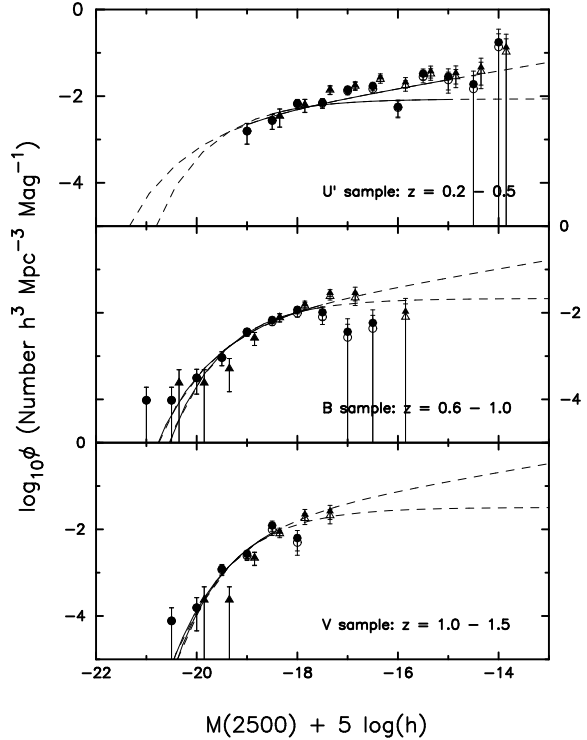}
\figcaption[paper_3plot_cosmo.ps]{
Circles as in Fig.~\ref{fig:3plotmethod} for a
$\Omega_{{\rm m}0} = 0.3, \Omega_{\lambda 0} = 0.7$ cosmology. Overlaid 
are the best-fitting Schechter functions 
assuming either $\alpha=-1.0$ or $\alpha=-1.5$.
The solid portion of the line shows the magnitude range utilized
in the fit.
Also shown and denoted by
the triangles (offset by 0.15~mag for clarity) are the 
minimal (open) and incompleteness-corrected (solid) functions 
obtained for an Einstein-de Sitter cosmology.
See text for details.
See Tables~{\ref{tab:bestfitFL}} and~{\ref{tab:bestfitEdS}} for
best Schechter function fit parameters.
\label{fig:3plotcosmo}
}
\end{figure}

We then investigated how one's choice of cosmology affects the LFs. 
We reconstructed 
LFs again using the $V_{\rm max}(M)$ method but this time assuming 
an Einstein-de Sitter cosmology. 
For comparison, 
in Fig.~\ref{fig:3plotcosmo}, the 
open triangles show the minimal function; 
the solid triangles the incompleteness-corrected function 
assuming this
cosmology. As in Figure~\ref{fig:3plotmethod},
the triangles have been offset by 0.15 mag for clarity. 

Table~{\ref{tab:bestfitEdS}} shows the best-fit Schechter values of absolute magnitude
at the knee ($M^\star$) and normalization ($\phi^\star$) for 
the Einstein-de Sitter cosmology (again assuming fixed faint-end slopes of
$-1.0$ and $-1.5$). From Figure~\ref{fig:3plotcosmo}
and from Tables~{\ref{tab:bestfitFL}} and~{\ref{tab:bestfitEdS}}, it is clear
that the larger 
distance/volumes associated with a cosmological constant cause 
$M^\star$ and $\phi^\star$ to decrease compared to the
best-fit parameters in the Einstein-de Sitter case.

%
%

\section{REST-FRAME UV LUMINOSITY DENSITY EVOLUTION}
\label{sec:ld}

The LFs in \S~\ref{sec:UVLF} can now be used to calculate the rest-frame ultraviolet 
luminosity densities, ${\mathcal L}$, with redshift. One approach is to choose a 
magnitude limit and to sum the LF over the magnitude bins directly using

\begin{equation}
\label{eq:Mtol}
{\mathcal L}_{\rm Direct} = 4.4 \times 10^{20} 
\sum\frac{10^{-0.4M}}{V_{\rm max}(M)} 
{h\ {\rm erg\ s}^{-1}\ {\rm Hz}^{-1}\ {\rm Mpc}^{-3}}
\end{equation}

\noindent
An alternative approach, which we adopt, is to choose a faint-end slope
(we use either $\alpha=-1.0$ or $\alpha=-1.5$) and to integrate the LF
analytically using the best-fit Schechter parameters from 
Tables~\ref{tab:bestfitFL} and~\ref{tab:bestfitEdS} 

\begin{equation}
\label{eq:l}
{\mathcal L}_{\rm Schechter}=\int^{\infty}_{0} L\phi(L) dL=
{L_{\star}} {\phi_{\star}} \Gamma(\alpha+2) 
\end{equation}

\noindent
giving

\begin{equation}
\label{eq:lphys}
{\mathcal L}_{\rm Schechter}  = 4.4 \times 10^{20} \times 10^{-0.4~M_{\star}} 
\phi_{\star} \Gamma(\alpha+2)~h
\end{equation}

\noindent 
in units of \ergsHzMpccube\ .
Although this method involves integrating
over all luminosities, fainter galaxies have a rapidly decreasing 
contribution to the total luminosity density, and thus the two methods 
give similar results. We calculate the luminosity density for both the minimal and incompleteness-corrected cases 
assuming firstly a faint-end slope of $\alpha = -1.0$ and then  $\alpha = -1.5$.  The resulting
luminosity densities for the flat lambda and Einstein de-Sitter cosmologies are shown in Table~\ref{tab:uvld}.

\begin{deluxetable}{ccccccc}
\tablewidth{0pt}
\tablecaption{Comoving 2500\AA\ UV Luminosity Density.\tablenotemark{a} 
\label{tab:uvld}
}
\tablewidth{0pt}
\tablehead{
\colhead{}  & \colhead{}  & \colhead{}  & \multicolumn{4}{c}{log$_{10}$(Luminosity Density) }\\
\colhead{}  & \colhead{}  & \colhead{}  & \multicolumn{2}{c}{Flat Lambda} & \multicolumn{2}{c}{Einstein-de Sitter}\\
\colhead{Sample}  & \colhead{Redshift} & \colhead{Corrected?} &  \colhead{$\alpha = -1.0$} & \colhead{$\alpha = -1.5$} &  \colhead{$\alpha = -1.0$} & \colhead{$\alpha = -1.5$}\\
}
\startdata
$U'$  &  $0.35\pm0.15$ & No  & $26.058$ & $26.211$ & $26.202$ & $26.340$ \
\\
$U'$  &  $0.35\pm0.15$ & Yes & $26.077$ & $26.234$ & $26.223$ & $26.360$ \
\\
$B$   &  $0.80\pm0.20$ & No  & $26.272$ & $26.420$ & $26.491$ & $26.681$  \
\\
$B$   &  $0.80\pm0.20$ & Yes & $26.339$ & $26.500$ & $26.546$ & $26.746$  \
\\
$V$   &  $1.35\pm0.25$ & No  & $26.284$ & $26.561$ & $26.600$ & $26.871$  \
\\
$V$   &  $1.35\pm0.25$ & No  & $26.428$ & $26.712$ & $26.695$ & $26.968$  \
\\
\enddata
\tablenotetext{a}{Units are $h$\ergsHzMpccube}
\end{deluxetable}

\begin{figure}
\centering\epsfig{file=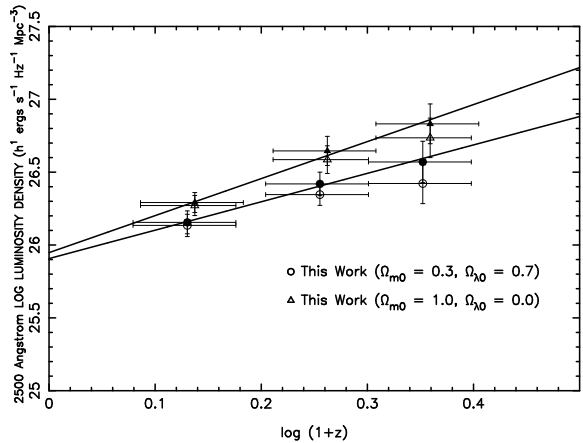,width=\figwidth}
\figcaption[plotlumdens_cosmo_loglog_inf_offset.ps]{
$\log_{10}$(Luminosity density) versus $\log_{10}(1+z)$. The open circles 
denote the minimal case and the solid circles 
denote the incompleteness-corrected case, both for the flat lambda cosmology.
The vertical bars show the uncertainties in luminosity density caused
by assuming $\alpha = -1.0$ (lower) or $\alpha=-1.5$ (upper).
The triangles show same and are for an Einstein de-Sitter cosmology.
They have been offset slightly for clarity.
Also shown are the best-fit power-law to the incompleteness-corrected
values for each cosmology.
See \S~\ref{sec:ld} for a discussion of the best 
power-law fits and Table~\ref{tab:bestpower} for the values.
\label{fig:lumdens_cosmo}
}
\end{figure}

Luminosity density evolution with redshift is often parameterized as a 
power-law, ${\mathcal L} \propto (1+z)^{\beta}$. In Fig.~\ref{fig:lumdens_cosmo}, 
we show $\log_{10}$(luminosity density) versus $\log_{10}(1+z)$ using the 
values from Table~\ref{tab:uvld}. As in Fig.~\ref{fig:3plotcosmo}, we use 
circles to denote the the flat 
lambda cosmology and triangles to denote the Einstein-de Sitter 
cosmology. The open symbols denote the 
minimal 
case , and the solid symbols denote the
incompleteness-corrected case.

%
%

\begin{deluxetable}{ccc}
\tablewidth{0pt}
\tablecaption{Best-fit Power Law Exponent $\beta$ 
\label{tab:bestpower}
}
\tablewidth{0pt}
\tablehead{
\colhead{Corrected?}  & \colhead{Flat Lambda} & \colhead{Einstein-de Sitter}
}
\startdata
No  & $1.44\pm0.63$ & $2.22\pm0.62$\
\\
Yes & $1.95\pm0.65$ & $2.54\pm0.62$\
\\
\enddata
\end{deluxetable}

We solved for the best-fit power-law exponent, $\beta$, in each case. 
We used the mean of the luminosity densities obtained assuming faint end slopes of
$\alpha = -1.0$ and $\alpha=-1.5$ as our best
estimate, with the extreme values as estimates of the uncertainty. 
Table~\ref{tab:bestpower} gives the best-fit exponent and uncertainty 
as a function of completeness-correction and cosmology. 
For the flat lambda cosmology
we found a best-fit exponent of $1.44\pm0.63$ in the minimal case and a
best-fit exponent of $1.95\pm0.65$ in the incompleteness-corrected case. 
For the Einstein de-Sitter cosmology we found a  best-fit exponent of 
$2.22\pm0.62$ in the minimal case and a best-fit exponent of $2.54\pm0.62$ 
in the incompleteness-corrected case. Thus, depending on the choice of 
completeness correction, we conclude that luminosity density evolves 
as $(1+z)^{1.7\pm1.0}$ in the $\Omega_{{\rm m}0} = 0.3, \Omega_{\lambda 0} = 0.7$ 
cosmology and as $(1+z)^{2.4\pm1.0}$ in the Einstein de-Sitter cosmology.
The two solid lines overlaid on Fig.~\ref{fig:lumdens_cosmo} show the 
best-fit solutions for each cosmology in the incompleteness-corrected case.  
The Einstein de-Sitter value is slightly steeper than that obtained by CSB 
($1.3$ for $\alpha = -1.0$ and $1.7$ for $\alpha = -1.5$) but consistent within 
the uncertainties.


%
%
\begin{figure}
\centering\epsfig{file=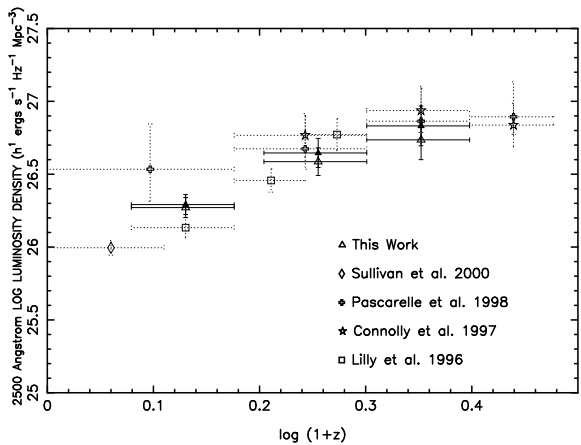,width=\figwidth}
\figcaption[plotlumdens_eds_loglog_inf.ps]{
$\log_{10}$(Luminosity density) versus $\log_{10}(1+z)$ for an 
Einstein de-Sitter cosmology. The triangles are as in 
Fig.~\ref{fig:lumdens_cosmo} and
denote the minimal (open) and incompleteness-corrected (solid) 
case. The vertical bars show the uncertainties in luminosity density caused
by assuming $\alpha = -1.0$ or $\alpha=-1.5$.
For comparison we have included data points from
Lilly {et~al.} (1996, open squares),
Connolly {et~al.} (1997, open star),
Pascarelle {et~al.} (1998, open cross) and
Sullivan {et~al.} (2000, open diamond).
\label{fig:lumdens_eds}
}
\end{figure}

Finally, in Fig.~\ref{fig:lumdens_eds}
we compare our values of luminosity density to the
values obtained by other surveys. Other groups have previously assumed 
an Einstein-de Sitter cosmology and therefore should be compared 
to the triangles from Fig.~\ref{fig:lumdens_cosmo}.
To convert the low-redshift value obtained by \markcite{sull-00}{Sullivan} {et~al.} (2000) 
(from the FOCA2000 balloon-born survey) we used their 
best-fit Schechter values (their Table 3) 
and converted their magnitudes to AB magnitudes using a 2.29~mag offset; 
we also converted from a rest-frame of 2000 \ang\ to 2500 \ang\ 
using a $\lambda^{1.1}$ power-law, as suggested 
by Fig.~4 of CSB. As mentioned in \S~\ref{sec:intro},  the \markcite{sull-00}{Sullivan} {et~al.} 
UV-selected sample
results in  a higher value of integrated luminosity density for the local universe
than previous optically-derived estimates.

In comparing with the surveys of \markcite{pas-98}Pascarelle, Lanzetta, \&  Fern$\acute{a}$ndez-Soto (1998), \markcite{con-97}Connolly {et~al.} (1997), and
\markcite{lilly-96}Lilly {et~al.} (1996), we again converted to a rest-frame of 2500~\ang\ using a $\lambda^{1.1}$ 
power-law. We also converted to $H_0$ = 100~$h$~km~s$^{-1}$~Mpc$^{-1}$, 
where necessary. \markcite{con-97}Connolly {et~al.} and 
\markcite{pas-98}Pascarelle {et~al.} both calculate 
their luminosity density from the small Hubble Deep Field North-proper 
(HDF-N)
using photometric redshift estimates. The results of 
\markcite{con-97}Connolly {et~al.} are more directly 
comparable to ours since they measure at a rest-frame of 2800 \AA\, 
whereas \markcite{pas-98}Pascarelle {et~al.} measure at a 
rest-frame of 1500 \AA . \markcite{con-97}Connolly {et~al.}  
assume a faint end slope of $\alpha=-1.3$. 
Both obtain somewhat higher luminosity densities than do we, although
the values are consistent within the uncertainties.
As discussed at length in the Appendix to CSB, much of the 
discrepancy between our results and \markcite{con-97}Connolly {et~al.} and 
\markcite{pas-98}Pascarelle {et~al.} may be attributable to the slightly 
higher number counts in the HDF-N versus other fields.

\markcite{lilly-96}Lilly {et~al.} calculated their luminosity density 
from the Canada-France Redshift Survey, and 
found a best-fit exponent of 
$\beta = 3.9\pm0.75$. As discussed in \S~\ref{sec:intro},
this is somewhat steeper then the value of $\beta = 2.5\pm1.0$ that we obtained. 
From Fig.~\ref{fig:lumdens_eds},
we conclude that the steeper value obtained by  \markcite{lilly-96}Lilly {et~al.} is most
likely due to a combination of their $z\sim1$ luminsity density estimate
being rather higher than ours, and their use of 
a low (optically-derived)
estimate of the local luminsity density.

In closing, it is important to add one caveat concerning the effect of
interstellar dust on our conclusions. In this paper we assumed
that any extinction would suppress UV emission uniformly. 
This corresponds to applying a constant correction
factor to the LFs, and does not affect the luminosity density
slope inferred from Fig.~\ref{fig:lumdens_eds}.
Some authors have suggested that extinction
may be luminosity dependent \markcite{adel-00,sull-00,hop-01}({Adelberger} \& {Steidel} 2000; {Sullivan} {et~al.} 2000; {Hopkins} {et~al.} 2001).
If this is the case, the higher redshift luminosity functions
containing greater contributions from brighter galaxies would see
larger extinction corrections, possibly flattening the slope observed in 
Fig.~\ref{fig:lumdens_eds} from a steeper value.
The satisfactory resolution of the complex role of dust
and the validity of these claims 
will require further investigation with larger samples.

\section{SUMMARY}
\label{sec:conc}

We investigated the evolution of the universal rest-frame luminosity 
density from $z = 1.5$ to the present. The availability of both multicolor 
data and highly complete spectroscopy enabled us to select galaxies 
based on their rest-frame ultraviolet color, minimizing potential 
sources of error such as large $K$-corrections and interstellar dust. 
Our large, deep sample allowed us to constrain the faint-end of the 
luminosity function with confidence, even at the highest redshift 
interval of $z = 1.25\pm0.25$. Assuming analytic Schechter forms for 
our luminosity functions and using likely extremal faint-end slope 
values of $\alpha = -1.0$ and $\alpha = -1.5$, we constrained the 
{\em relative} luminosity density as a function of redshift.  We 
concluded that, in an $\Omega_{{\rm m}0} = 0.3, \Omega_{\lambda 0} = 0.7$ 
[Einstein-de Sitter] Universe, the evolution in the luminosity 
density follows a $(1+z)^{1.7\pm1.0}$ [$(1+z)^{2.4\pm1.0}$] 
slope from $z = 1.5$ to the present, implying that rather more star 
formation has occurred in recent times than was previously suggested.

\acknowledgements 
We gratefully acknowledge support from NASA through
Hubble Fellowship grant HF-01117.01-A (AJB) awarded by the
Space Telescope Science Institute, which is operated by the
Association of Universities for Research in Astronomy, Inc.,
for NASA under contract NAS 5-26555, the
University of Wisconsin Research Committee with funds
granted by the Wisconsin Alumni Research Foundation (AJB),
NSF grants AST-0084847 (AJB, PI) and AST-0084816 (LLC),
SAO contract SV4-64008 (DJB) and NASA contract NAS8-39073 (DJB).

\clearpage




\end{document}